\documentstyle[jltp,11pt,epsf,multicol]{article}
\epsfverbosetrue
\newcommand{\onlinecite}{\citen}
\newcommand{\be}{\begin{equation}}
\newcommand{\ee}{\end{equation}}
\newcommand{\grad}{\mbox{\boldmath$\nabla$}}

\newcommand{\tinyonehalf}{\frac{\mbox{\tiny 1}}{\mbox{\tiny 2}}}

\def\vp{{\bf p}}
\def\vq{{\bf q}}
\def\vR{{\bf R}}
\def\vv{{\bf v}}
\def\lsim{{\buildrel <\over \sim}}
\title{
\vspace*{-40mm}{\normalsize\sl
Proc. of QFS97 ({\it Paris})- to appear in Journ. of Low Temp. Phys.}\\
\vspace*{15mm}
Sound Propagation and Transport Properties of Liquid $^3$He in Aerogel
}
\author{D.\ Rainer$^{\mbox{\tiny 1}}$ and J.\ A.\ Sauls$^{\mbox{\tiny 2}}$}
\address{
  $^{\mbox{\tiny 1}}$Physikalisches Institut,
  Universit\"at Bayreuth,
  Bayreuth, Germany\\[1ex]
  $^{\mbox{\tiny 2}}$Department of Physics \& Astronomy,
  Northwestern University,
  Evanston, Illinois\\[1ex]
  {\rm (Received August 31, 1997)}
}
\runninghead{D. Rainer and J. A. Sauls}{Sound Propagation
and Transport Properties of Liquid $^3$He in Aerogel}
\begin{document}
%
\maketitle
\vspace{-1.5cm}
\centerline{\bf Abstract}
\noindent{\it Superfluid $^3$He confined in aerogel offers
a unique chance to study the
effects of a short mean free path on the properties of a
well defined superfluid Fermi liquid with anisotropic pairing.
Transport coefficients and 
collective excitations, e.g. longitudinal sound, are expected
to react sensitively to a short mean free path and 
to offer the possibility for testing
recently developed models for quasiparticle
scattering at aerogel strands. Sound experiments, together with
a theoretical analysis based on Fermi liquid theory for
systems with short mean free paths, should give valuable insights
into the interaction between superfluid $^3$He and aerogel.}
%
\vspace{0.3in}
%

A model for liquid $^3$He in aerogel based on
a random distribution of short-ranged potentials 
acting on $^3$He quasiparticles has been shown
to account semi-quantitatively for the reduction
of the transition temperature, $T_c$,
and the suppression of the superfluid density,
$\rho_s(T)$.\cite{thu96}
Although the order parameter for $^3$He in aerogel 
is not firmly identified,
measurements of the magnetization indicate
that the low pressure phase is an ESP state.\cite{spr95}
A transverse NMR
shift is observed and is roughly consistent with
that of an axial state even though the B-phase is stable
in pure $^3$He at these pressures. 
The addition of a small concentration of $^4$He, which coats the 
aerogel strands, induces a suppression of the magnetization for 
$T<T_c^{\mbox{\tiny aerogel}}$, indicative of a non-ESP state like
the BW phase.\cite{spr95} Thus, the stability of the superfluid
state of $^3$He in aerogel is quite sensitive to the detailed interaction
between $^3$He and the aerogel strands. Calculations of the free
energy for $^3$He in aerogel
which include anisotropic and magnetic scattering, and the effects
of orientational correlations
of the aerogel strands, confirm the sensitivity of the superfluid phases
to the interaction between $^3$He and the aerogel.\cite{thu96,sau96}

The high porosity of the aerogel implies that the silica structure
does not significantly modify the bulk properties of normal $^3$He.
The dominant effect of the aerogel structure is
to scatter $^3$He quasiparticles moving at the bulk Fermi velocity.
If the coherence length, $\xi_0=\hbar v_f/2\pi T_{c0}$,
is sufficiently long compared to the average distance
between scattering centers
then a reasonable starting point is to treat
the aerogel as a homogenous scattering
medium (HSM) described by a mean-free path $\ell$.\cite{thu96}
In this article we discuss the effects of a short mean free path
on some of the transport properties of liquid $^3$He. 
The calculations presented below for sound propagation
assume the BW phase is stable; however, the propagation and damping
of low-frequency sound are expected to be qualitatively similar 
for other phases.

\bigskip
\centerline{\epsfxsize=0.75\textwidth\epsfbox{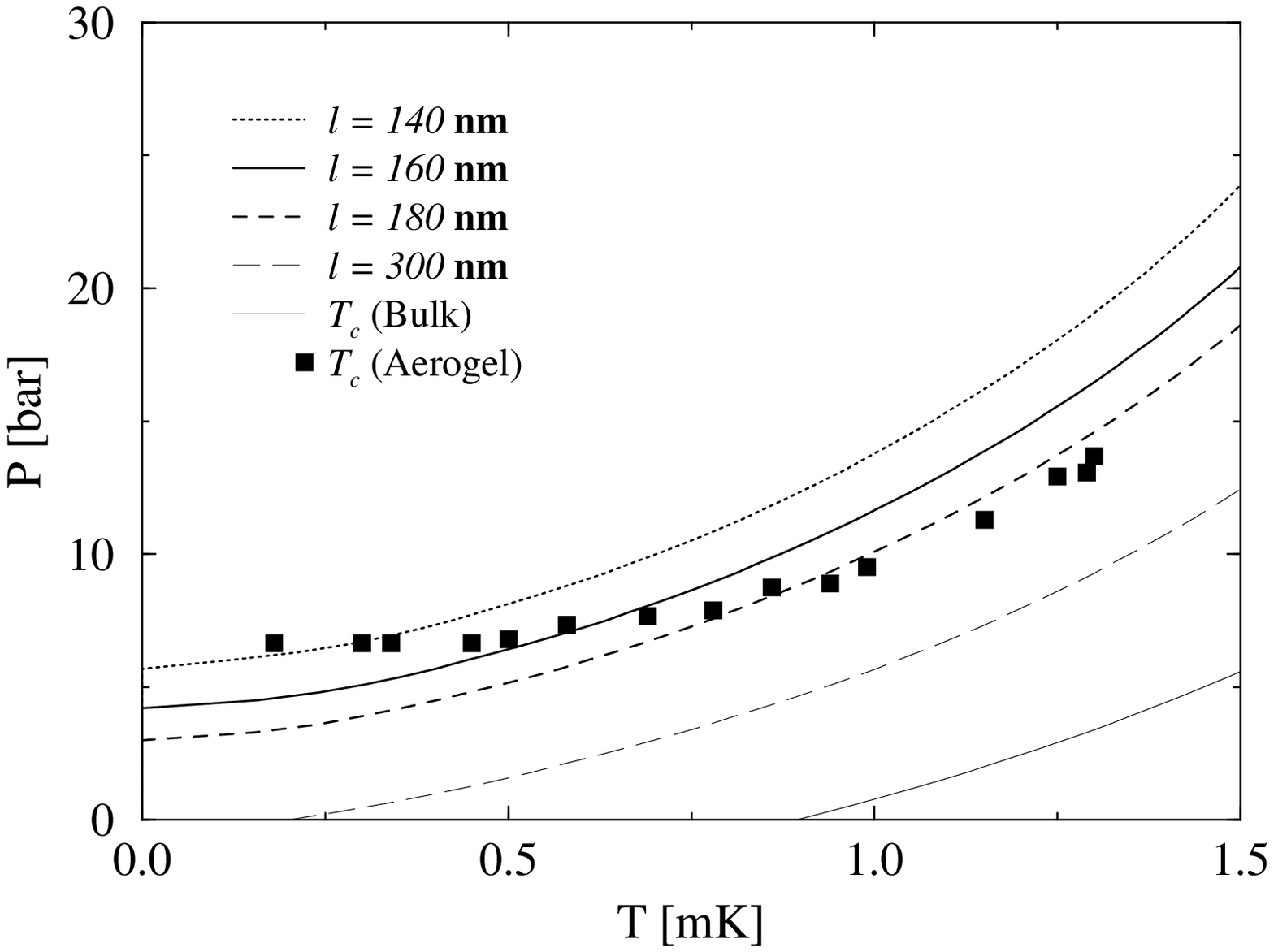}}
\begin{center}
\begin{minipage}{0.8\textwidth}
Fig. 1 - $p$-$T$ phase diagram vs. $\ell$. The 
geometric mean-free path
is estimated to be $1750\,\AA$. The
data is from Ref. \onlinecite{mat97}.
\end{minipage}
\end{center}

The superfluid transition temperature for $^3$He in aerogel
is suppressed by quasiparticle scattering off the
aerogel structure. In the HSM model the suppression
of $T_c$ is given by the Abrikosov-Gorkov formula,
\be\begin{array}{c}
\ln(T_{c0}/T_{c})=
\Psi\left(\tinyonehalf+\tinyonehalf\frac{\xi_0}{\ell}\frac{T_{c0}}{T_c}\right)
-\Psi\left(\tinyonehalf\right)
\,,
\end{array}
\ee
with the pair-breaking parameter given by $\xi_0/\ell$.\cite{thu96}
Figure 1 shows the calculated aerogel transition temperature
for several values of the mean-free path.
The pressure dependence of the calculated
phase diagram is determined by the pressure dependence of
the bulk transition temperature and the Fermi velocity via
$\xi_0=\hbar v_f/2\pi T_{c0}$. The experimental data for $T_c(p)$
is from Matsumoto, et al.\cite{mat97}
for an aerogel with $\approx 98\%$ open volume.
For this porosity the typical
diameter of the aerogel strands is $d\simeq 30\,\AA$ and the
mean distance between strands is $D\simeq 325\,\AA$, which
should be compared with the bulk coherence length
$\xi_0\simeq 700\,\AA$ at $p=1\,{\rm bar}$.
The geometric mean-free path of the aerogel 
determined from the surface to volume
ratio, $S\simeq 2.6\times 10^{5}\,\mbox{cm}^{-1}$, is
$\ell_{\mbox{\tiny geom}}
=\frac{3\pi}{2}/S\simeq 1750\,\AA$.
The calculations show a mean-free-path of this order gives
good agreement with the phase diagram at low temperatures and
low pressures, i.e. for
$\xi(p)=\hbar v_f/2\pi k_B T_c(p,\ell) > D\simeq 325\,\AA$,
which corresponds to pressures $p\lsim 15\,\mbox{bar}$.
Also note that the mean-field phase diagram shows
a zero-temperature phase transition as a function of
pressure determined by $\xi_0(p_c)=0.28\,\ell$.

The transport properties of $^3$He should also be strongly
affected by scattering from the aerogel. In the normal state
the quasiparticle distribution function,
$\phi(\hat{\vp},\vR;\epsilon,t)$, satisfies the
Boltzmann-Landau transport equation,
\be
\frac{\partial\phi}{\partial t}+\vv_f\cdot\grad\phi
+\left(\frac{\partial\phi_0}{\partial\epsilon}\right)
\frac{\partial{\cal E}}{\partial t}=I[\phi]
\,,
\ee
where ${\cal E}$ is the effective potential acting on
the quasiparticles and $I[\phi]$ is the collision integral.
The effective potential consists of the external driving 
potentials and the internal potentials resulting from
quasiparticle interactions.
In pure liquid $^3$He the quasiparticle lifetime is
determined by inelastic collisions between quasiparticles
and is of order $\tau_{\mbox{\tiny in}}
\simeq 1\mu \mbox{s-mK}^2/T^2$ at $p=15\,\mbox{bar}$.
However, in aerogel the 
mean time between scatterings by the aerogel is
$\tau_{\mbox{\tiny el}}=\ell/v_f\simeq 4\,\mbox{ns}$ at the
same pressure. Thus, for temperatures below 
$T_*\simeq 16\,\mbox{mK}$
the collision rate is dominated by quasiparticle
scattering by the aerogel: $I_{el}=-\frac{1}{\tau_{\tiny el}}
\left(\phi - \left<\phi\right>_{\mbox{\tiny FS}}\right)$.
This leads to strong reduction
in the thermal conductivity and viscosity, and to 
an increase in the damping of hydrodynamic sound. 
The static transport coefficients exhibit cross-over
behavior dictated by the scattering rate;
the thermal conductivity and shear viscosity
scale as
\be
\kappa=\frac{\mbox{\small 1}}{\mbox{\small 3}}
C_v v_f^2\,\tau\sim
\Bigg\{
	\begin{array}{l}
        1/T \\
         T
	\end{array}
\,\, , \,\,
\eta=
\frac{\mbox{\small 1}}{\mbox{\small 15}} N_fv_f^2p_f^2\,\tau\sim
\Bigg\{
        \begin{array}{l}
        1/T^2          \quad,\quad T>T_* \\
        \mbox{\small const}\quad,\quad T<T_* 
        \end{array}
\,,
\ee
above and below $T_*$. Recent speculations that the
low-temperature phase of $^3$He in aerogel for $p<p_{cr}$
is not described by Fermi-liquid theory\cite{mat97,par97}
can be tested by measuring these transport coefficients.

At low frequencies a compressible aerogel
will move nearly in phase with the $^3$He density
and longitudinal current mode; sound propagates but it is
damped by the viscous coupling of the $^3$He to the
aerogel, $\alpha_1=\frac{\omega^2}{\rho c_1^3}\eta$,
where $c_1$ is the hydrodyanmic sound velocity and $\rho$
is the mass density of $^3$He.
The viscous damping of hydrodynamic sound
saturates for $T<T^*$ at $\alpha_1/q\simeq
\frac{\mbox{\tiny 3}}{\mbox{\tiny 5}}(\omega\tau_{el})/(1+F^s_0)$.
At higher frequencies the impedance mismatch
between $^3$He and the aerogel sound mode leads to an increasingly
out-of-phase motion of $^3$He excitations and aerogel.
Hydrodynamic sound may become overdamped  and reemerge
as a diffusive mode. The frequency at which the
cross-over from damped hydrodynamic
sound to an overdamped diffusive mode occurs depends 
on the elastic compliance of the aerogel and the microscopic
details of the coupling between $^3$He and the aerogel strands.
Here we assume the frequency is
above this cross-over, in which case the hydrodynamic 
mode is an overdamped diffusive mode, $\omega=-i{\cal D}_s\,q^2$,
where ${\cal D}_s=c_1^2\,\tau$ is the acoustic
diffusion constant with $1/\tau=(1+F^s_1/3)/\tau_{el}$.
At still higher frequencies, $\omega\tau\gg 1$, zero sound
can propagate, albeit with relatively high attenuation,
$\alpha_0\simeq1/c_0\tau\sim 10^{4}\,\mbox{cm}^{-1}$ at 
$p=10\,\mbox{bar}$. The regime for propagating zero sound 
is also pushed to higher frequencies,
$\omega/2\pi >v_f/2\pi\ell\simeq 42\,\mbox{MHz}$
at $p=10\,\mbox{bar}$.

\bigskip
\centerline{\epsfxsize=0.75\textwidth\epsfysize=0.55\textwidth
\epsfbox{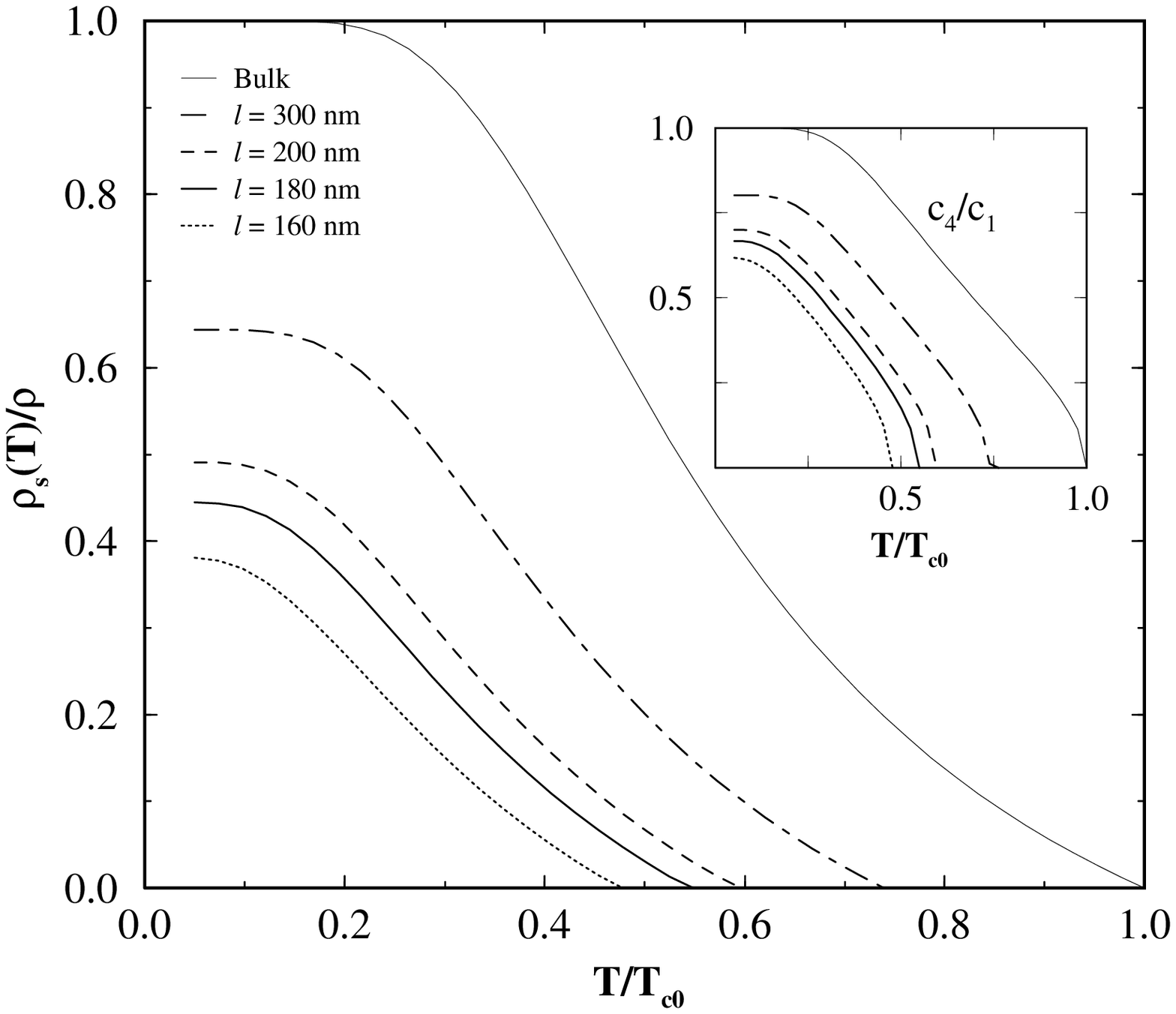}}
\begin{center}
\begin{minipage}{0.8\textwidth}
Fig. 2 - Suppression of $\rho_s(T)$ by scattering 
at $p=10\,\mbox{bar}$ with $F_1^s=8.6$. 
Results in Born approximation 
 are shown for bulk $^3$He-B and mean-free paths
$\ell=1600$ to $3000\,\AA$. The inset shows
the $4^{\mbox{\tiny th}}$ sound velocity for
the same values of the mean-free path.
\end{minipage}
\end{center}

The collective mode spectrum of superfluid $^3$He in aerogel is
also expected to show significant changes compared to bulk $^3$He.
\cite{mckenna91,mulders91,kopf97} A typical example is low frequency
sound in superfluid $^3$He
($\omega\ll \Delta/\hbar$). In bulk $^3$He one has collisionless zero sound 
and hydrodynamic first sound with 
nearly the same velocities, $c^2_0\approx c^2_1={1\over 3}(1+F^s_0)
(1+{1\over 3}F_1^s)v_f^2$, and small damping by 
quasiparticle-quasiparticle scattering
(for reviews on sound and collective modes in $^3$He
see e.g. Refs. 
\onlinecite{wol78,hal90,mck90}). 
In aerogel, on the other hand, one
expects a behavior which is similar to sound in $^3$He
confined to small channels.\cite{koj74} 
Sound will be weakly damped for $q\ell\gg 1$, ceases to be a well defined 
mode for $q\ell\approx 1$, and reappears for $q\ell\ll 1$ 
as fourth sound 
with a temperature dependent velocity,
$c_4\simeq\sqrt{\rho_s(T)/\rho}\,c_1$.
Fig. 2 shows the effects of elastic scattering
on the superfluid density.\cite{cho88} 
Note the reduction in $\rho_s/\rho$ at $T=0$. For a 
mean-free path of $\ell=1800\,\AA$ less than
$50\,\%$ of the $^3$He
mass density contributes to $\rho_s(T=0)$. The 
inset shows the fourth sound
velocity neglecting damping.

\begin{center}
\begin{minipage}{0.8\textwidth}
\noindent{\epsfxsize=\textwidth\epsfbox{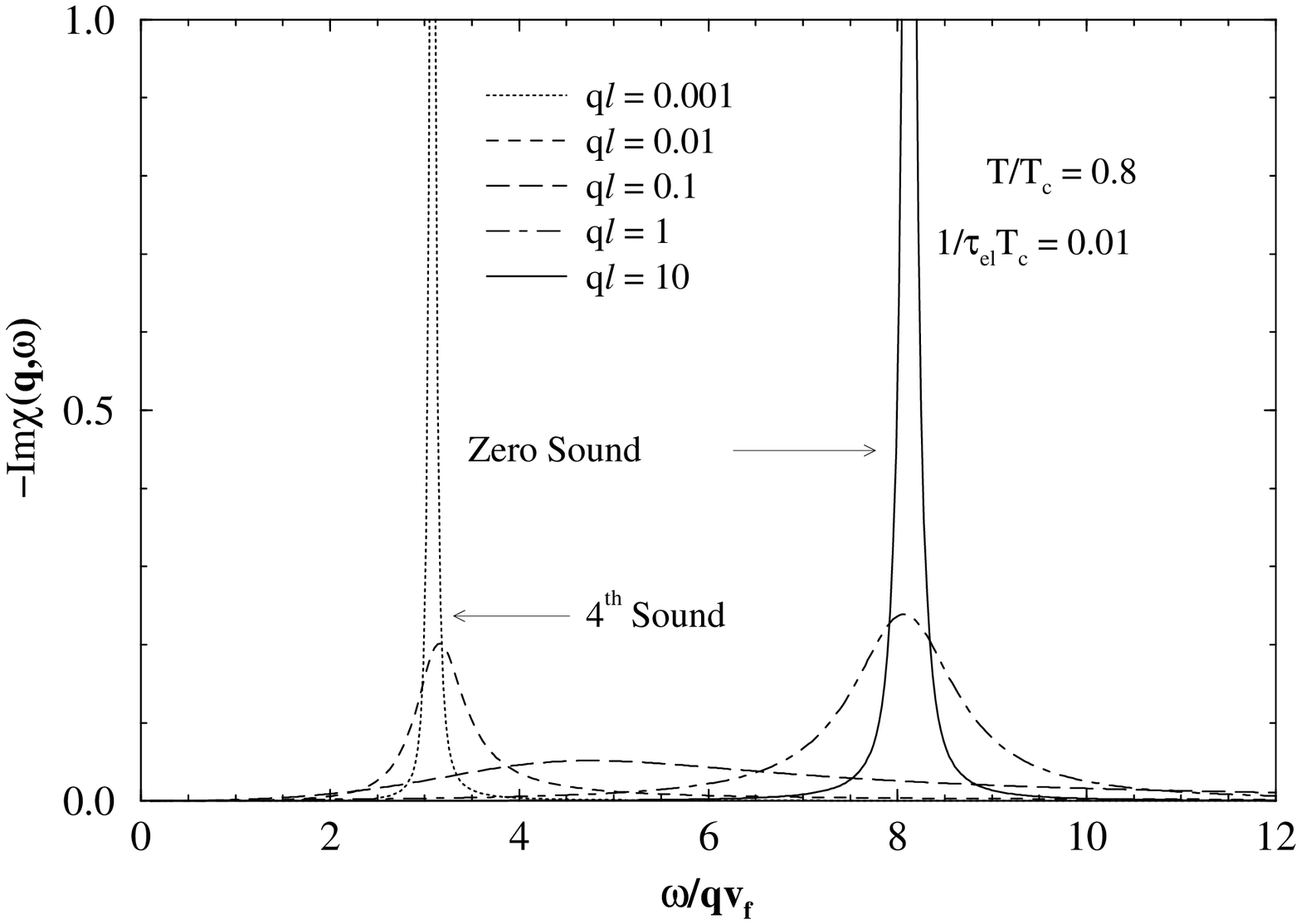}}
\parindent=-5mm
\begin{center}
\begin{minipage}{0.95\hsize}
Fig. 3 - Spectral function of the 
density response $\chi(\vq,\omega)$ for $q\ell=
0.001\, ...\, 10$. The results are for 
$T/T_c=0.8$, $1/\tau_{el}T_c = 0.01$,
$F_0^s=50$ and $F_1^s=8.6$; the corresponding zero sound
mode is at $\omega/v_fq = c_0/v_f = 8.1$.
\end{minipage}
\end{center}
\end{minipage}
\end{center}

Measurements of sound propagation in $^3$He-aerogel
should provide a sensitive test of the HSM model.
We study the spectrum of longitudinal sound in this model by
calculating the linear 
response of  the $^3$He density, $\rho({\bf q};\omega)$, to a driving force of 
wavevector ${\bf q}$ and frequency $\omega$. Our driving force will be a
scalar field described by a potential, $\delta U_{ext}({\bf q};\omega)$,  
which couples to the $^3$He density. 
More realistic ``experimental 
driving forces'' require  more elaborate calculations, which
are not called for given the present experimental status.
The calculation
follows the quasiclassical version\cite{ser83}
of the method of Betbeder-Matibet and Nozi\`eres.\cite{bet69}
In the low frequency limit one has to solve
Boltzmann-Landau transport equations for
the branches of particle-like and hole-like excitations with
distribution functions 
$\delta\phi_{B1,B2}(\hat\vp,{\bf R};\epsilon,t)$.
 We keep
the dominant Landau parameters, $F_0^s$ and $F_1^s$,  and 
obtain an effective  scalar potential, 
$\delta\tilde u({\bf q};\omega)$, and longitudinal
vector potential, 
$\delta\tilde a({\bf q};\omega)={\bf v}_f\cdot
\delta{\bf A}({\bf q};\omega)$.
The collision terms in the HMS model have the form,
$I_{B1,B2}=-{1 \over\tau(\epsilon)}\left( 
\delta\phi_{B1,B2}
 -\Big\langle\delta\phi_{B1,B2}\Big\rangle_{FS}\right)$,
where $\tau(\epsilon)=\ell/v(\epsilon)$, and
$v(\epsilon)=v_f\sqrt{\epsilon^2- \mid\Delta\mid^2}/\epsilon$
is the energy dependent quasiparticle velocity in the superfluid state.
In the low frequency limit the transport equations have to be supplemented
by Landau's self-consistency equations for the effective potentials,
$\delta\tilde u$ and $\delta\tilde a$, and the particle conservation law,
$\dot{\rho}+\grad\cdot{\bf j}=0$. 
In this limit we can ignore the less important 
self-consistency equation for the amplitude, $\delta\mid\Delta\mid$,
of the order parameter. The five coupled 
linear equations for the distribution functions, the
effective potentials, and the 
phase, $\delta\Psi({\bf q};\omega)$, of the order parameter 
can be solved, and will be described elsewhere.

\begin{center}
\begin{minipage}{0.8\textwidth}
\parindent=-5mm
\centerline{\epsfxsize=\textwidth\epsfbox{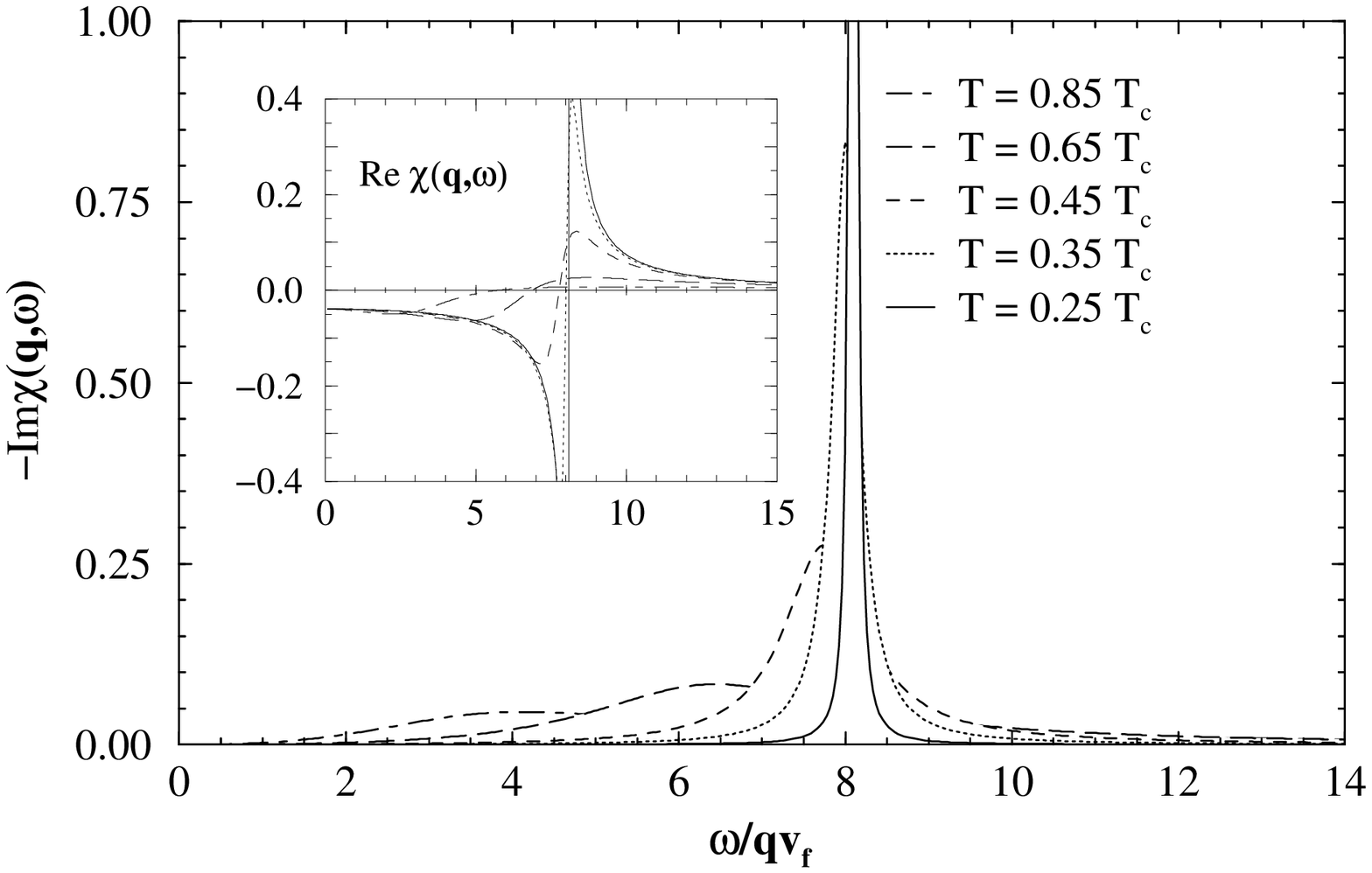}}
\begin{center}
\begin{minipage}{0.95\hsize}
Fig. 4 - Temperature dependence of the imaginary
and real (inset) parts of the spectral function $\chi(\vq,\omega)$
for $T/Tc = 0.25$ ... $0.85$.
The results are shown for $q\ell=0.1$, $qv_f
=0.001T_c$,
$F_0^s=50$ and $F_1^s=8.6$.
\end{minipage}
\end{center}
\end{minipage}
\end{center}

The calculated crossover from weakly damped zero
sound to weakly damped fourth sound
is shown in Fig. 3. We display the frequency dependent spectral function,
$-{\cal I}\mbox{\small m}\chi(q,\omega)$, for various wavelengths at fixed 
temperature and elastic scattering time.
One can see clearly the transition from zero sound at 
$v_fq\gg 1/\tau_{el}$ to fourth sound at $v_fq\ll 1/\tau_{el}$. 
Fig. 4 shows the real and imaginary parts 
of the response function, $\chi(q,\omega)$, at various temperatures 
and  fixed $\tau_{el}$.
Because of the short mean free path, $q\ell=0.1$,
the zero sound resonance with wavevector $\bf{q}$
is overdamped in the normal
state and just below $T_c$. The damping
decreases exponentially in the superfluid state for
$T\ll\Delta$ because of the freezing out
of thermally excited quasiparticles; and, 
as can be seen from Fig. 4, a well defined zero
sound mode develops.

\newpage

We thank
the Alexander von Humboldt-Stiftung, the Deutsche Forschungsgemeinschaft
(SFB279) and the STC for Superconductivity (NSF 91-20000)
for their support.

\end{document}